\documentclass[10pt, showpacs,preprintnumbers]{revtex4}
\usepackage{graphicx}
\usepackage{floatrow}
\usepackage[caption=false]{subfig}
\usepackage{xcolor}
\usepackage{amsmath}
\usepackage{amssymb}
\usepackage{mleftright}
\usepackage{dsfont}
\mleftright
\def\bc{\begin{center}}
\def\ec{\end{center}}

\def\beq{\begin{equation}}
\def\eeq{\end{equation}}

\begin{document}

\title{Laplace transform approach for the dynamics of $N$ qubits coupled to a resonator}
\author{Mirko Amico, Oleg L. Berman and Roman Ya. Kezerashvili}
\affiliation{\mbox{Physics Department, New York City College
of Technology, The City University of New York,} \\
Brooklyn, NY 11201, USA \\
\mbox{The Graduate School and University Center, The
City University of New York,} \\
New York, NY 10016, USA}

\begin{abstract}

An approach to use the method of Laplace transform for the perturbative solution of the Schr\"{o}dinger equation at any order of the perturbation 
for a system of $N$ qubits coupled to a cavity with $n$ photons is suggested. We investigate the dynamics of a system of $N$ superconducting qubits 
coupled to a common resonator with time-dependent coupling. To account for the contribution of the dynamical Lamb effect to the probability of excitation 
of the qubit, we consider counter-rotating terms in the qubit-photon interaction Hamiltonian. As an example, we illustrate the method for the case of two qubits coupled to a common cavity. 
The perturbative solutions for the probability of excitation of the qubit show excellent agreement with the numerical calculations.

\end{abstract}
\pacs{03.65.Ud, 03.67.Bg, 42.50.Dv, 42.50.Ct, 85.25.Am}

\maketitle


In this article we investigate the dynamics of a system of $N$ superconducting qubits coupled to a common resonator with time-dependent coupling using 
the method of Laplace transform. The technique of Laplace transform is widely used in the study of electrical circuits \cite{decarlo}. It is interesting to note that the Laplace transform has been previously used to study the dynamics of a superconducting qubit coupled to a resonator \cite{tureci1, tureci2}. Here, we extend its application and successfully apply the Laplace transform to the study of the phenomenology of superconducting circuits. If the coupling is periodic in time, the method of Laplace transform can be used to solve the Schr\"{o}dinger equation. In facts, the Laplace transform turns a differential equations in the time domain into an algebraic one in the Laplace domain, allowing to easily find a solution of Schr\"{o}dinger equation. The strength of the method lies in the possibility of using complex analysis for the calculation of the inverse Laplace transformation which transforms back to the time domain. This reduces to the application of the Cauchy residue theorem and, therefore, to the calculation of the \textit{Residues}.

For weak qubit/cavity coupling, the Schr\"{o}dinger equation describing the dynamics of the system can be treated in a perturbative approach.
A perturbative analytical solution of the Schr\"{o}dinger equation obtained in Ref. \cite{amico2} shows excellent agreement with numerical calculations when one considers a time-averaged coupling. However, the approximation of constant coupling only allows to probe a limited range of frequency of switching of the coupling $\varpi_s$, as it is valid only for high frequency of switching $\varpi_s \gg 2 \omega_0$, where $\omega_0$ is the transition frequency of the qubit. We extend these results by going beyond the constant coupling approximation and considering the dynamics of the system in the case of time-dependent coupling. 
We obtain a perturbative solution to the Schr\"{o}dinger equation for a system of $N$ qubit coupled to a cavity where $n$ photons are present in the Laplace domain. The inverse Laplace transformation then allows to transform the solution to the time domain. 

Following Ref. \cite{lozovik}, we go beyond the rotating-wave approximation and consider counter-rotating terms in the Hamiltonian of qubit-photon interaction which give rise to a new quantum vacuum phenomenon, the dynamical Lamb effect (DLE).
The DLE is the parametric excitation of an atom, along with the creation of photons from the vacuum, due to the instantaneous change in its Lamb shift \cite{lamb, bethe}.
The DLE was first predicted in Ref. \cite{lozovik} and further studied in Refs. \cite{shapiro, zhukov, remizov, berman, amico1, amico2}. Different physical realization with superconducting qubits coupled to a coplanar waveguide have been proposed for the observation of the DLE. For example, in Refs. \cite{berman, amico1} a system of two and three superconducting qubits coupled to a common nonstationary cavity is considered. The nonadiabatic change in effective length of the cavity can be achieved by terminating the coplanar waveguide with a superconducting quantum interference device, thus giving rise to the DLE. In Refs. \cite{shapiro, zhukov, remizov, amico2} a different approach which would make the DLE the dominant source of excitation of the qubit is adopted. It consists of periodically switching on/off the qubit/cavity coupling nonadiabatically. In facts, one can mimic the sudden change in Lamb shift which happens to an atom passing from a cavity to another at relativistic speed by the instantaneous switching on/off of the coupling between a superconducting qubit and a coplanar waveguide playing the role of a cavity. Periodical switching increases the probability of excitation of the qubit and the creation of photons dramatically \cite{zhukov}.
Several aspects related to the DLE have been previously studied. For example, the probability of excitation of the qubit and the creation of photons were investigated in Refs. \cite{lozovik, shapiro, berman, amico1}, the entanglement generated by the DLE in Refs. \cite{remizov, berman, amico1, amico2} and the effects of dissipation were considered in Refs. \cite{zhukov, remizov, amico2}.
 
To illustrate the effectiveness of our approach we apply the procedure for the case of two qubits coupled to a common cavity. The probability of excitation of the qubits is calculated using the Laplace method within a perturbative approach and compared to numerical calculations. Excellent agreement is found for a broad range of values of the frequency of switching. 


Let us consider the Schr\"{o}dinger equation

\begin{equation}
\label{schroedinger}
i \frac{d {\lvert \psi \left( t \right)  \rangle} }{dt} = \hat{H}(t) {\lvert \psi \left( t \right)  \rangle} .
\end{equation}

\noindent
In Eq. (\ref{schroedinger}) $H(t)$ is the Hamiltonian of the system and the wavefunction is written in terms of the $N$ qubits and $n$ photons as

\begin{equation}
\label{psi}
\lvert \psi \left( t \right)  \rangle  = \sum_{i=0}^{n} \alpha_{gg...g, i} \left( t \right) \lvert gg...g , i   \rangle +  \alpha_{ge...g , i} \left( t \right) \lvert ge...g , i   \rangle + ... + \alpha_{ee...e , i } \left( t \right) \lvert ee...e , i   \rangle ,
\end{equation}

\noindent
where $\alpha (t)$ are the time-dependent coefficients which describe the time evolution of the corresponding states, $i$ counts the number of photons in the cavity and $g,e$ represents a qubit in the ground or excited state, respectively.
The Laplace transform of the wavefunction (\ref{psi}) is
\begin{equation}
\label{Psi}
\lvert \Psi \left( s \right)  \rangle  = \sum_{i=0}^{n} A_{gg...g, i} \left( s \right) \lvert gg...g , i   \rangle +  A_{ge...g , i} \left( s \right) \lvert ge...g , i   \rangle + ... + A_{ee...e , i } \left( s \right) \lvert ee...e , i   \rangle ,
\end{equation}

\noindent
where

\begin{equation}
 A (s) = \mathcal{L}\left[ \alpha (t) \right](s) = \int_0^{\infty} dt \, \alpha (t) e^{-st} 
\end{equation}
 
\noindent 
is the Laplace transform of $\alpha (t)$.

To model a system of $N$ superconducting qubits coupled to a single-mode resonator we use the Tavis-Cummings Hamiltonian \cite{tavis} 

\begin{equation}
\label{H}
\hat{H}\left( t \right) = \hat{H}_{0} + \delta \hat{H}_{I}\left( t \right) ,
\end{equation}

\noindent
where $\hat{H}_{0}$ is the Hamiltonian of the qubits and the cavity mode, $\hat{H}_{I}\left( t \right)$ is the time-dependent Hamiltonian which accounts for the qubit/cavity interaction and $\delta$ is a small dimensionless parameter which is used to define the qubit/cavity coupling.

The Hamiltonian of the non-interacting system ($\hbar = 1$) is

\begin{equation}
\label{H0}
\hat{H}_{0}  = \omega_{c} \hat{a}^{\dagger} \hat{a} + \omega_0 \sum_{i=1}^{N} \hat{\sigma}_{i}^{+} \hat{\sigma}_{i}^{-} ,
\end{equation}

\noindent
where $\hat{a}^{\dagger}$ and $\hat{a}$ are the creation and annihilation operators for the cavity photons and $\hat{\sigma}^{-} = \frac{\hat{\sigma}_1 -i \hat{\sigma}_2}{2}$, $\hat{\sigma}^{+} = \frac{\hat{\sigma}_1 +i \hat{\sigma}_2}{2}$ are defined via the Pauli matrices $\hat{\sigma}_1$ and $\hat{\sigma}_2$ for each qubit.
The Hamiltonian of the qubit-photon interaction $ \hat{H}_{I}\left( t \right)$ is split in the following way

\begin{equation}
\label{HI}
 \hat{H}_{I}\left( t \right) = \hat{V}_{\text{RWA}}(t)+\hat{V}(t).
\end{equation}

\noindent
Here, $\hat{V}_{\text{RWA}}(t) =  g \left( t \right) \sum_{i=1}^{N}\left(\hat{\sigma}_{i}^{+}a + \hat{\sigma}_{i}^{-}a^{\dagger}\right)$ includes the qubit-photon interaction in the rotating wave approximation (RWA), which does
not change the number of the excitations in the system, and $\hat{V}(t)= g \left( t \right)
\sum_{i=1}^{N} \left( \hat{\sigma}_{i}^{+}a^{\dagger }+\hat{\sigma}_{i}^{-}a \right) $ contains terms beyond the RWA, which do not conserve the number of the excitations in the system.
To account for the DLE, we go beyond the rotating wave approximation and consider both $ \hat{V}_{RWA}(t)$ and $\hat{V}(t)$ as a perturbation to $\hat{H}_{0}$.
$\delta g \left( t \right)$ is the time-dependent qubit/cavity coupling, and we are taking $g(t)$ as

\begin{equation}
\label{g}
g \left( t \right) = g_0 \theta \left( \cos \varpi_s t \right),
\end{equation}

\noindent
where $\theta \left( \cdot \right)$ is the Heaviside function. Let us note that $\delta g_0$ is the qubit/cavity coupling strength. The expression in Eq. (\ref{g}) represents a square wave signal that switches on/off periodically between $0$ and $g_0$ with a period $T_s=\frac{2 \pi}{\varpi_s}$. Equivalently, this can be written in a form which is more suitable for analytical calculations as 

\noindent
\begin{equation}
\label{g2}
g(t)=g_0 \left\{ \frac{1}{2} + \frac{1}{2} \sum_{k=0}^{\infty} \left[ \theta \left( t - kT_s \right) -2 \theta \left( t - \frac{2k+1}{2}T_s \right) + \theta \left( t - (k+1)T_s \right)   \right] \right\} .
\end{equation}

\noindent
In the Laplace domain, the time-dependent qubit/cavity coupling becomes

\begin{equation}
\label{G2}
G(s)= \mathcal{L} \left[ g(t) \right] =  \frac{g_0}{2s} \left\{ 1+ \sum_{k=0}^{\infty} \left[ e^{- ksT_s s} -2 e^{- \frac{2k+1}{2}T_s s} + e^{- (k+1)T_s s}\right]  \right\} = \frac{g_0}{2s} \frac{1}{1+e^{-\frac{T_s s}{2} } } \, .
\end{equation}

\noindent
The latter expression allows to find the Laplace transform of the Hamiltonian (\ref{H})

\begin{equation}
\label{Hs}
\hat{H}\left( s \right) = \omega_{c} \hat{a}^{\dagger} \hat{a} + \omega_0 \sum_{i=1}^{N} \hat{\sigma}_{i}^{+} \hat{\sigma}_{i}^{-} + \delta G(s)  \sum_{i=1}^{N} \left( \hat{\sigma}_{i}^{+}a + \hat{\sigma}_{i}^{-}a^{\dagger} + \hat{\sigma}_{i}^{+}a^{\dagger }+\hat{\sigma}_{i}^{-}a \right)   .
\end{equation}

\noindent
We can now write the Schr\"{o}dinger equation in the Laplace domain as

\begin{equation}
\label{Schroedinger}
i s \lvert \Psi \left( s \right)  \rangle - \lvert \psi \left( 0 \right)  \rangle   = \hat{H}(s) {\lvert \Psi \left( s \right)  \rangle} .
\end{equation}

\noindent
Substituting the wavefunction (\ref{Psi}) and the Hamiltonian (\ref{Hs}) in Eq. (\ref{Schroedinger}), and imposing the initial condition $\lvert \psi \left( 0 \right)  \rangle = 1$, a recurrent algebraic equation for the coefficients $A(s)$ can be obtained

\begin{equation}
\begin{split}
\label{A}
i s A_{x_0x_1 \ldots x_N,n } (s) -1 = \left[ \omega_c n + \omega_0 \left( \bar{x} \cdot \bar{1} \right) \right] {A}_{x_0x_1 \ldots x_N,n }(s) +  \sum_{l=0}^{N} G(s) *  \left( \sqrt{n} \delta_{ x_{l} -1, 0} {A}_{x_0x_1 \ldots x_{l} -1 \dots x_N,n-1 }(s) + \right. \\
\left. + \sqrt{n+1} \delta_{ x_{l} +1, 1} {A}_{x_0x_1 \ldots x_{l} +1 \dots x_N,n+1 }(s) +  \sqrt{n} \delta_{ x_{l} +1, 0} {A}_{x_0x_1 \ldots x_{l} +1 \dots x_N,n-1 }(s) + \sqrt{n+1} \delta_{ x_{l} -1, 1} {A}_{x_0x_1 \ldots x_{l} -1 \dots x_N,n+1 }\left(s\right) \right) ,
\end{split}
\end{equation}

\noindent
where $\bar{x}$ stands for the $N$-bit string which represents the state of the qubits as a bit-string (zero for the ground state $g$ and one for the excited state $e$), $x_{l}$ denotes the $l$-th element of the $N$-bit string, $\bar{x} \cdot \bar{1}$ counts the number of qubit's excitations by taking the dot product between the $N$-bit string and the string of all ones, namely $\bar{x} \cdot \bar{1} = x_0 1 + x_1 1 + x_2 1 + ... + x_N 1$.
Also, $*$ is the convolution product of Laplace transforms, which is defined as $ F(s)*U(s) = \frac{1}{2\pi i} \int_{c-i\infty}^{c+i\infty} d\sigma \,  F(\sigma)U(s-\sigma)$, with $c$ a point on the real line on the right of the rightmost pole of the integrand.

The system of equations which can be obtained from Eq. (\ref{A}) cannot be easily solved because of the integral implicit in the convolution product.
However, if the qubit/cavity coupling strength $\delta g_0$ is smaller than the qubit transition frequency $\omega_0$ and the frequency of the cavity photons $\omega_c$, we can solve Eq. (\ref{A}) within a perturbative approach.
First, one can expand the wavefunctions (\ref{Psi}) and (\ref{psi}) in terms of $\delta$

\begin{equation}
\label{Psi_exp}
{\lvert \Psi \left( s \right)  \rangle} = {\lvert \Psi \left( s \right)  \rangle}^{(0)} +\delta {\lvert \Psi \left( s \right)  \rangle}^{(1)} + \delta^2 {\lvert \Psi \left( s \right)  \rangle}^{(2)} + ... \, ,
\end{equation}

\begin{equation}
\label{psi_exp}
{\lvert \psi \left( t \right)  \rangle} = {\lvert \psi \left( t \right)  \rangle}^{(0)} +\delta {\lvert \psi \left( t \right)  \rangle}^{(1)} + \delta^2 {\lvert \psi \left( t \right)  \rangle}^{(2)} + ... \, .
\end{equation}

\noindent
Then, one can solve Eq. (\ref{A}) order by order in the perturbation $\delta$

\begin{equation}
\begin{split}
\label{A_j}
i s A_{x_0x_1 \ldots x_N,n }^{(j)} (s) - \alpha_{x_0x_1 \ldots x_N,n }^{(j)} (t=0) = \left[ \omega_c n + \omega_0 \left( \bar{x} \cdot \bar{1} \right) \right] {A}_{x_0x_1 \ldots x_N,n }^{(j)}(s) +  \sum_{l=0}^{N} G(s) *  \left( \sqrt{n} \delta_{ x_{l} -1, 0} {A}_{x_0x_1 \ldots x_{l} -1 \dots x_N,n-1 }^{(j-1)}(s) + \right. \\
\left. + \sqrt{n+1} \delta_{ x_{l} +1, 1} {A}_{x_0x_1 \ldots x_{l} +1 \dots x_N,n+1 }^{(j-1)}(s) +  \sqrt{n} \delta_{ x_{l} +1, 0} {A}_{x_0x_1 \ldots x_{l} +1 \dots x_N,n-1 }^{(j-1)}(s) + \sqrt{n+1} \delta_{ x_{l} -1, 1} {A}_{x_0x_1 \ldots x_{l} -1 \dots x_N,n+1 }^{(j-1)}(s) \right) ,
\end{split}
\end{equation}

\noindent
where $A_{x_0x_1 \ldots x_N,n }^{(j)} (s)$ and $\alpha_{x_0x_1 \ldots x_N,n }^{(j)}(t)$ are the $j$-th order  coefficient, which are obtained by expanding the wavefunctions (\ref{Psi}) and (\ref{psi}) in terms of $\delta$. Assuming that the system is in the ground state at time $t=0$ gives the initial condition $\alpha_{0 0 \ldots 0,0 }^{(0)}(t=0)=1$.   
Eq. (\ref{A_j}) gives a set of coupled equation describing the dynamics of the system of $N$ qubits for a fixed amount of cavity photons $n$ in the Laplace domain at any order of the perturbation $(j)$. 
To find the final solution, one has to solve Eq. (\ref{A_j}) order by order and then transform back to the time-domain by taking the inverse Laplace transform. 
The latter is defined as 

\begin{equation}
\label{inv_laplace} 
\mathcal{L}^{-1}\left[ F(s) \right](t) = \frac{1}{2\pi i} \int_{b-i\infty}^{b+i\infty} ds \, e^{st} F(s) = \sum_{\text{poles of } F(s)} \text{Res}\left( F(s)e^{st} \right), 
\end{equation}

\noindent
where $b$ is again a point on the real line on the right of the rightmost pole of $F(s)$. Therefore, one can determine the perturbative time-dependent coefficients, thus the perturbative wavefunction.


\begin{figure}[t]
	{
		\label{20wa}
		{\includegraphics[width = 2.5 in]{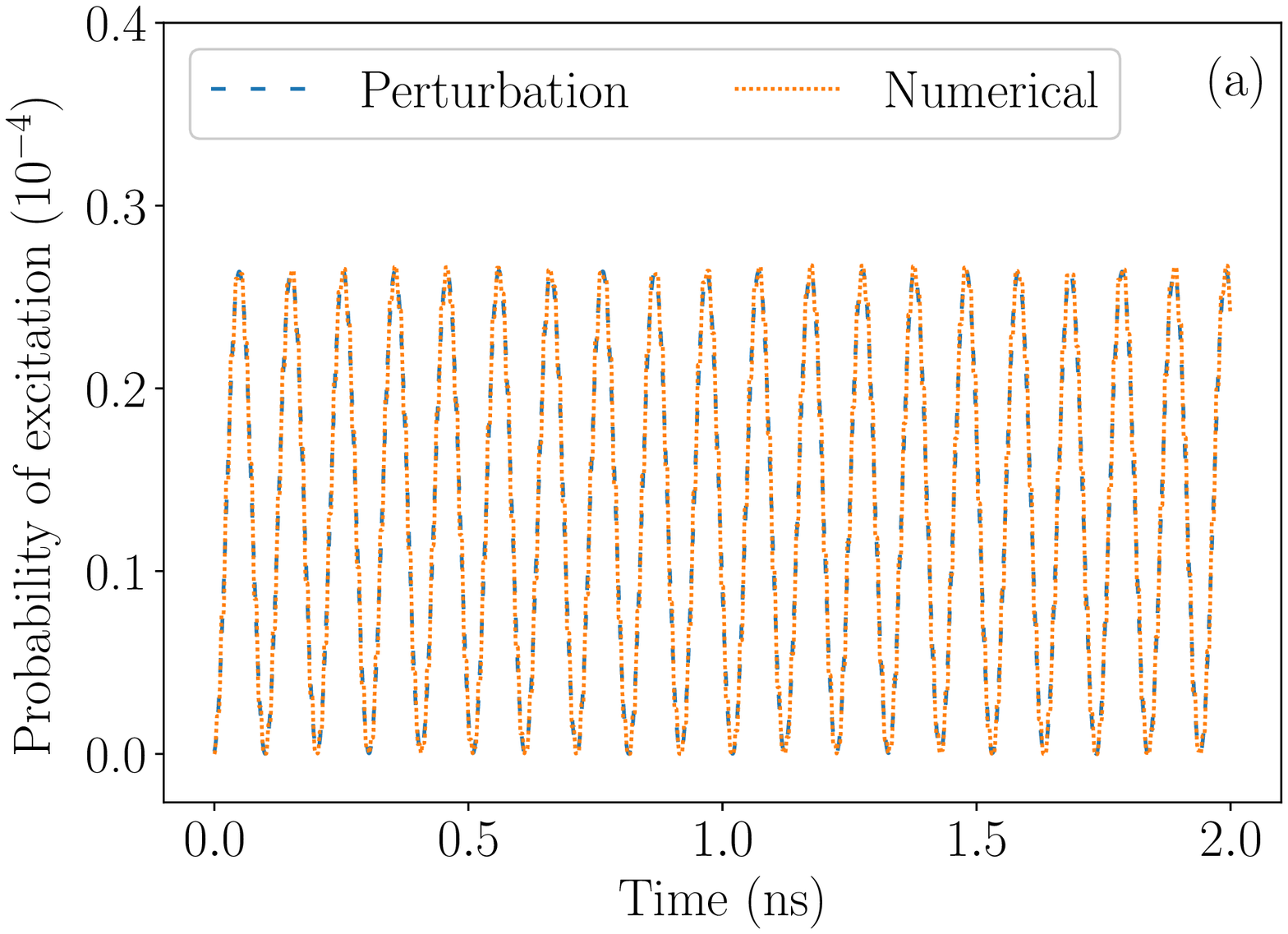}}
		}
	{
		\label{10wa}
		{\includegraphics[width = 2.5in]{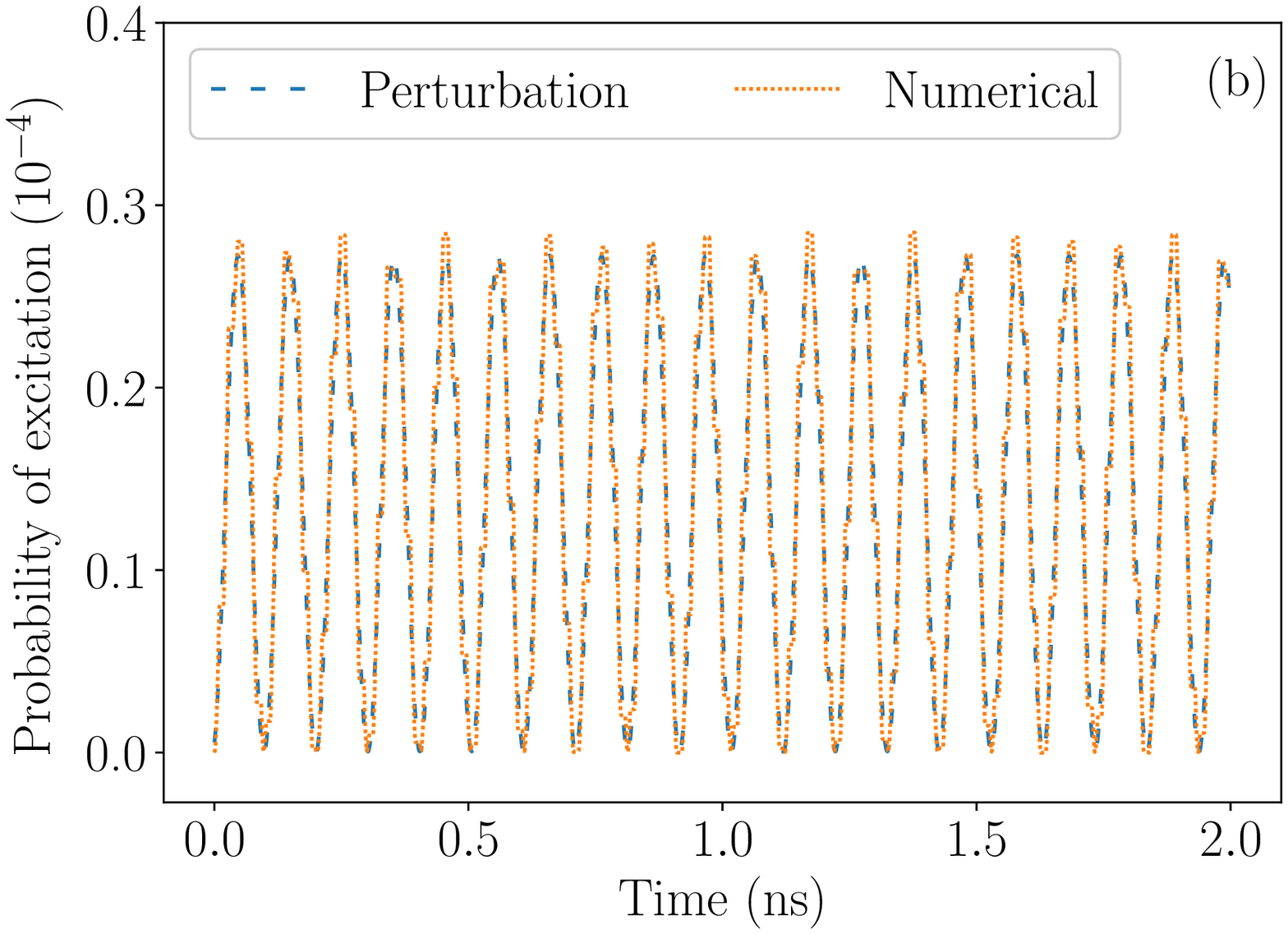}}
		} \\
	{
		\label{5wa}
		{\includegraphics[width = 2.5in]{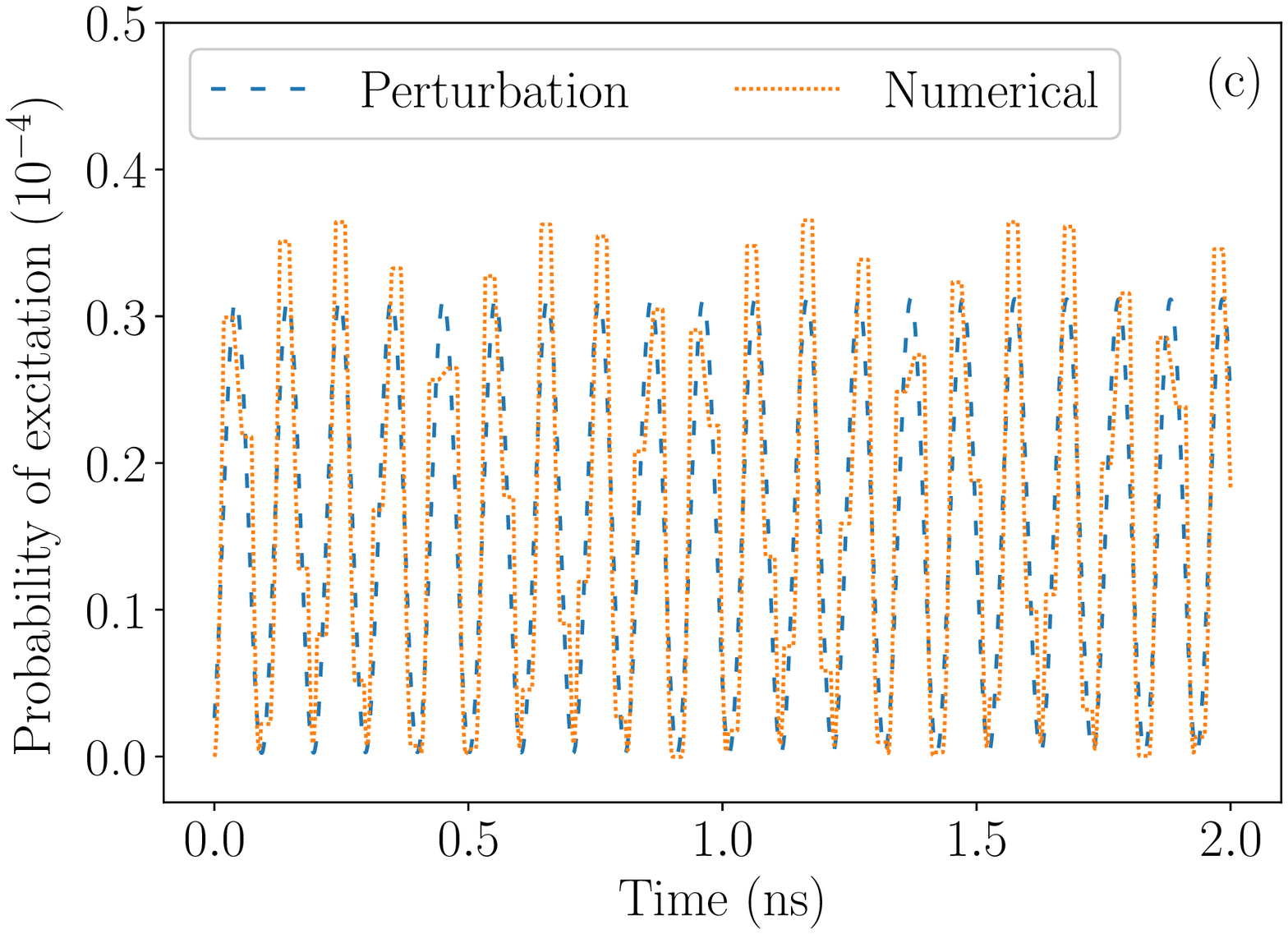}} 
		}
	{
		\label{2wa}
		{\includegraphics[width = 2.5in]{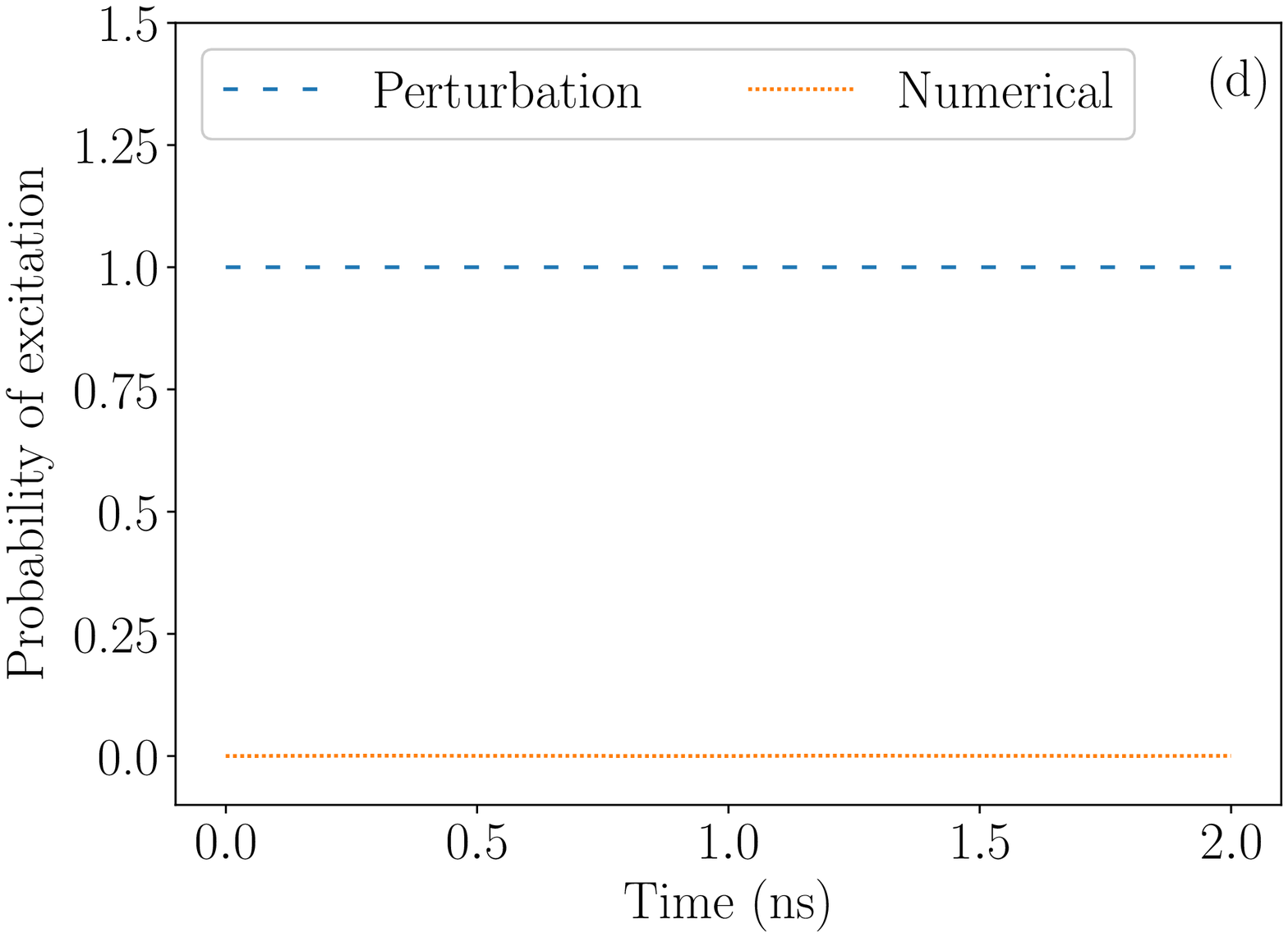}} 
		}
	\caption{The figures show the time evolution of the probability of excitation of one qubit in a system of two qubits and $n = 0,1$ photons for different values of the frequency of switching of the coupling $\varpi_s$. The perturbative calculations are compared with the numerical ones. Excellent agreement between is found in (a) and (b), corresponding to frequency of switching of the coupling of $\varpi_s =20 \omega_0$ and $\varpi_s =10 \omega_0$, respectively. For lower frequency of switching of the coupling, $\varpi_s =5 \omega_0$ in (c), the different calculations give slightly different results. (d) The perturbative approach breaks down at $\varpi_s  = 2 \omega_0$, where a unit probability of excitation of the qubit is obtained because the wavefunction parametrically diverges for this value of $\varpi_s$ (see Eq. (A15) in Appendix).  }
\label{p_ex}
\end{figure}

As an example, we illustrate the above procedure and provide the corresponding solutions for the case of $N=2$ qubits coupled to a common resonator with $n=0,1$ photons. However, the method developed is valid for any number of qubits and photons. To avoid lengthy mathematical expressions, we report here only the results obtained for this case. The details of the calculations can be found in the Appendix. Fig. \ref{p_ex}, shows a comparison between the perturbative calculations within the Laplace transform approach and the numerical ones for the probability of excitation of one of the qubits. We use typical values of the parameters of the system taken from experiment \cite{lu}. In particular, we take $\omega_0 = 2 \pi \times 5.439$ GHz, $\omega_c = 2 \pi \times 4.343$ GHz and $\delta g_0 = 2 \pi \times 50$ MHz. The results show excellent agreement between the perturbative calculations and the numerical ones for high frequency of switching of the coupling, $\varpi_s \gg 2 \omega_0$ and $\varpi_s \gg \omega_c + \omega_0$. Decreasing $\varpi_s$, the comparison worsen as one approaches one of the resonances of the system, $\varpi_s \rightarrow 2 \omega_0$, $\varpi_s \rightarrow  \omega_0 + \omega_c$ or $\varpi_s \rightarrow  \omega_c - \omega_0$. Until it becomes completely inaccurate when $\varpi_s$ reaches one of the latter values. For example, the case $\varpi_s = 2\omega_0$ is shown in Fig. 1d. 

For this value of $\varpi_s$, the second order contribution ${\lvert ee,0  \rangle}^{(2)}$ of the wavefunction obtained in Eq. (A15) parametrically diverges at $\varpi_s = 2 \omega_0$, giving a unit probability of excitation of the qubit. Thus, the perturbative wavefunction cannot be reliably used to calculate the probability of excitation of the qubits for this case. Similar considerations apply for the other values of $\varpi_s$ for which the perturbative wavefunction diverges, $\varpi_s =  \omega_0 + \omega_c$ and $\varpi_s =  \omega_c - \omega_0$. Nonetheless, the divergence in the perturbative wavefunction suggests that these value of the frequency of switching of the coupling might lead to interesting effects. Indeed, in a previous work \cite{amico2}, we found through numerical calculations that for these value of the frequency of switching of the coupling all quantities used to measure the entanglement between the qubits reach their maximum value periodically.

In conclusion, we develop an approach where the method of Laplace transform is used to obtain the perturbative solution of the Schr\"{o}dinger equation at any order of the perturbation for a system of $N$ qubits coupled to a cavity, where $n$ photons are present. Our approach provides a method to solve the Schr\"{o}dinger equation in the case of time-dependent coupling avoiding the approximation of time-averaged coupling. In particular, this allows to describe the dynamics of the system beyond the rotating-wave approximation and for periodic switching of the qubit/cavity coupling, which makes the dynamical Lamb effect the dominant source of excitation of the qubits.
As an example to illustrate the method, we considered a system of two qubits coupled to a common cavity with $n=0,1$ photons. 
A perturbative solution of the Schr\"{o}dinger equation for a periodically switching coupling is found with the method of Laplace transform. 
The analytical expression for the perturbative wavefunction obtained can be used to give an accurate description of the dynamics of the system for almost any value of $\varpi_s$. 
In particular, we calculate the probability of excitation of one of the qubits in the system due to the DLE. The perturbative calculation for 
the probability of excitation of the qubit show excellent agreement with 
the numerical ones for $\varpi_s \neq 2 \omega_0$, $\varpi_s \neq \omega_c + \omega_0$ and $\varpi_s \neq \omega_c - \omega_0$. 
However, the perturbative approach employed breaks down at particular values of the frequency of switching, which may hint to interesting physical effects. 
We do not consider the effects of dissipation, as in a previous study \cite{amico2} it was noticed that dissipative effects have a negligible 
influence on the system's dynamics for typical values of dissipation rates present in experiments. 
This can easily be understood by noting that we are interested in phenomena arising in the nonadiabatic regime, which take place at time scales much 
shorter than that of dissipative effects.

\appendix

\section{Two qubits coupled to a common resonator}

\label{appendixa}

To illustrate the approach presented in the article, let us consider a system of $N=2$ qubits coupled to a cavity where $n=0,1$ photons are present. Note that the method is valid for an arbitrary number of qubits and photons. We restrict ourself to a simple case to make the exposition clearer. The Schr\"{o}dinger equation 
describing the dynamics of the system in the Laplace domain can be rewritten using the perturbative approach as a set of coupled algebraic equations by following the recurrent procedure given by Eq. (16). For the case considered, at the zero-th order in terms of $\delta$, Eq. (16) gives

\begin{eqnarray}
\label{coeff2q00_app}
i s {A}^{(0)}_{gg,0}  - i = 0 ,\nonumber \\ 
i s {A}^{(0)}_{ge,0}  = \omega_0 {A}^{(0)}_{ge,0}  ,\nonumber \\
i s {A}^{(0)}_{eg,0}  = \omega_0 {A}^{(0)}_{eg,0} ,\nonumber \\
i s {A}^{(0)}_{ee,0}  = 2\omega_0 {A}^{(0)}_{ee,0}  ,
\end{eqnarray}

\begin{eqnarray}
\label{coeff2q01_app}
i s {A}^{(0)}_{gg,1}  = \omega_c {A}^{(0)}_{gg,1} , \nonumber \\ 
i s {A}^{(0)}_{ge,1}  = \left( \omega_c + \omega_0 \right) {A}^{(0)}_{ge,1} , \nonumber \\
i s {A}^{(0)}_{eg,1}  = \left( \omega_c + \omega_0 \right) {A}^{(0)}_{eg,1} , \nonumber \\
i s {A}^{(0)}_{ee,1}  = \left( \omega_c + 2\omega_0 \right) {A}^{(0)}_{ee,1} .
\end{eqnarray}

\noindent
The solution is trivial and it reads $A^{(0)}_{gg,0}(s) = \frac{1}{s}$ which gives $\alpha^{(0)}_{gg,0}(t) = 1$.

At first order in terms of $\delta$, we have 

\begin{eqnarray}
\label{coeff2q10_app}
i s {A}^{(1)}_{gg,0}  =   G(s) * \left( {A}^{(0)}_{ge,1} + {A}^{(0)}_{eg,1} \right)   ,\nonumber \\ 
i s {A}^{(1)}_{ge,0}  = \omega_0 {A}^{(1)}_{ge,0} +  G(s) *  \left( {A}^{(0)}_{gg,1} + {A}^{(0)}_{ee,1} \right)   ,\nonumber \\
i s {A}^{(1)}_{eg,0}  = \omega_0 {A}^{(1)}_{eg,0} + G(s) *  \left( {A}^{(0)}_{gg,1} + {A}^{(0)}_{ee,1} \right)  ,\nonumber \\
i s {A}^{(1)}_{ee,0}  =  2\omega_0 {A}^{(1)}_{ee,0} +  G(s) *  \left( {A}^{(0)}_{ge,1} + {A}^{(0)}_{eg,1} \right)  ,
\end{eqnarray}

\begin{eqnarray}
\label{coeff2q11_app}
i s {A}^{(1)}_{gg,1}  = \omega_c{A}^{(1)}_{gg,1} +  G(s) *  \left (  {A}^{(0)}_{ge,0} + {A}^{(0)}_{eg,0} \right)   ,\nonumber \\ 
i s {A}^{(1)}_{ge,1}  = \left( \omega_c + \omega_0 \right) {A}^{(1)}_{ge,1} +  G(s) * \left( {A}^{(0)}_{gg,0} + {A}^{(0)}_{ee,0} \right)    ,\nonumber \\
i s {A}^{(1)}_{eg,1}  = \left( \omega_c + \omega_0 \right) {A}^{(1)}_{eg,1} +    G(s) * \left( {A}^{(0)}_{gg,0} + {A}^{(0)}_{ee,0} \right)  ,\nonumber \\
i s {A}^{(1)}_{ee,1}  =  \left( \omega_c + 2\omega_0 \right) {A}^{(1)}_{ee,1} + G(s) *  \left( {A}^{(0)}_{ge,0} + {A}^{(0)}_{eg,0} \right)  .
\end{eqnarray}

\noindent
The only equations which give a non-vanishing solutions are the second and third equations in (\ref{coeff2q11_app}). Which give

\begin{equation}
\label{y1s}
A^{(1)}_{1,ge}(s) = -\frac{i g_0}{s \left(e^{-\frac{s T_s}{2}}+1\right) (s+i(\omega_0+\omega_c))} ,
\end{equation}

\begin{equation}
\label{y1s2}
A^{(1)}_{1,eg}(s) = -\frac{i g_0}{s \left(e^{-\frac{s T_s}{2}}+1\right) (s+i(\omega_0+\omega_c))} ,
\end{equation}

\noindent
Once we found the solutions (\ref{y1s}) and (\ref{y1s2}) in Laplace space, we can take the inverse Laplace transform (16) to obtain their expressions in the time domain

\begin{equation}
\label{y1t}
\alpha^{(1)}_{1,ge}(t) = \frac{g_0} {2 (\omega_0+\omega_c)}\left(-1+\frac{2 e^{-i t
   (\omega_0+\omega_c)}}{1+e^{\frac{1}{2} i T_s
   (\omega_0+\omega_c)}}\right)
\end{equation}

\begin{equation}
\label{y1t2}
\alpha^{(1)}_{1,eg}(t) = \frac{g_0} {2 (\omega_0+\omega_c)}\left(-1+\frac{2 e^{-i t
   (\omega_0+\omega_c)}}{1+e^{\frac{1}{2} i T_s
   (\omega_0+\omega_c)}}\right)
\end{equation}

\noindent
which is the equivalent of Eq. (A5) in Ref. \cite{amico2} for the general case of switching of the coupling at any frequency. It can be easily checked that the two solutions coincide in the limit of high frequency of switching of the coupling $\varpi_s \rightarrow \infty$ or $T_s \rightarrow 0$.

Let us consider the set of equations obtained at second order in terms of $\delta$

\begin{eqnarray}
\label{coeff2q20_app}
i s{A}^{(2)}_{gg,0}  = G(s) *  \left(   {A}^{(1)}_{ge,1} + {A}^{(1)}_{eg,1} \right) ,\nonumber \\ 
i s {A}^{(2)}_{ge,0}  = \omega_0 {A}^{(2)}_{ge,0} +   G(s) *  \left(  {A}^{(1)}_{gg,1} + {A}^{(1)}_{ee,1}  \right)   ,\nonumber \\
i s {A}^{(2)}_{eg,0}  = \omega_0 {A}^{(2)}_{eg,0} +   G(s) *  \left( {A}^{(1)}_{gg,1} + {A}^{(1)}_{ee,1} \right)   ,\nonumber \\
i s {A}^{(2)}_{ee,0}  =  2\omega_0 {A}^{(2)}_{ee,0} +  G(s) *  \left(  {A}^{(1)}_{ge,1} + {A}^{(1)}_{eg,1} \right)    ,
\end{eqnarray}

\begin{eqnarray}
\label{coeff2q21_app}
i s {A}^{(2)}_{gg,1}  =  \omega_c{A}^{(2)}_{gg,1} +   G(s) *  \left(  {A}^{(1)}_{ge,0} + {A}^{(1)}_{eg,0}  \right)  ,\nonumber \\ 
i s {A}^{(2)}_{ge,1}  = \left( \omega_c + \omega_0 \right) {A}^{(2)}_{ge,1} +  G(s) *  \left(   {A}^{(1)}_{gg,0} + {A}^{(1)}_{ee,0}  \right)   ,\nonumber \\
i s {A}^{(2)}_{eg,1}  = \left( \omega_c + \omega_0 \right) {A}^{(2)}_{eg,1} +  G(s) *  \left(   {A}^{(1)}_{gg,0} + {A}^{(1)}_{ee,0}  \right)  ,\nonumber \\
i s {A}^{(2)}_{ee,1}  =  \left( \omega_c + 2\omega_0 \right) {A}^{(2)}_{ee,1} +  G(s) * \left(   {A}^{(1)}_{ge,0} + {A}^{(1)}_{eg,0}  \right)   .
\end{eqnarray}

\noindent
In this case, the only equations which do not give a vanishing solution are the first and the fourth equations in (\ref{coeff2q20_app}).
The solution is

\begin{align}
\label{y2s}
A_{0,gg}^{(2)}(s) = -\frac{i g_0^2 \left(-\tanh \left(\frac{s T_s}{4}\right) (s+2 i
   (\omega_0+\omega_c))+\left(s-i s \tan \left(\frac{1}{4} T_s
   (\omega_0+\omega_c)\right)\right) \tanh \left(\frac{1}{4} T_s
   (s+i (\omega_0+\omega_c))\right)-i
   (\omega_0+\omega_c)\right)}{2 s^2 (\omega_0+\omega_c) (s+i
   (\omega_0+\omega_c))} ,
\end{align}

\begin{equation}
\begin{split}
\label{y3s}
A_{0,ee}^{(2)}(s) = -\frac{i g_0^2 \left(-\tanh \left(\frac{s T_s}{4}\right) (s+2 i
   (\omega_0+\omega_c))+\left(s-i s \tan \left(\frac{1}{4} T_s
   (\omega_0+\omega_c)\right)\right) \tanh \left(\frac{1}{4} T_s
   (s+i (\omega_0+\omega_c))\right)-i
   (\omega_0+\omega_c)\right)}{2 s (s+2 i \omega_0)
   (\omega_0+\omega_c) (s+i (\omega_0+\omega_c))} .
   \end{split}
	\end{equation}

\noindent
The corresponding solutions in the time-domain are then found by applying the inverse Laplace transform (16):

\begin{equation}
	\begin{split}
	\label{y2t}
		\alpha_{0,gg}^{(2)} \left(t\right) = \frac{g_0^2 \left(i (2 t+T_s) (\omega_0+\omega_c)-2 i
   \left(1+e^{-i t (\omega_0+\omega_c)}\right) \tan
   \left(\frac{1}{4} T_s (\omega_0+\omega_c)\right)+2 e^{-i t
   (\omega_0+\omega_c)}-2 \sec ^2\left(\frac{1}{4} T_s
   (\omega_0+\omega_c)\right)\right)}{4 (\omega_0+\omega_c)^2}  ,	 
   \end{split}
\end{equation}

\begin{equation}
	\begin{split}
	\label{y3t}
	\alpha_{0,ee}^{(2)}\left(t\right) = \frac{1}{4} g_0^2 \left(\frac{2 i e^{-i t
   (\omega_0+\omega_c)} \left(\tan \left(\frac{1}{4} T_s
   (\omega_0+\omega_c)\right)+i\right)}{\omega_0^2-\omega_c^2} + \right. \\  
   \left. +\frac{e^{-2 i t \omega_0} \left(2 \omega_0 \tan
   \left(\frac{1}{4} T_s (\omega_c-\omega_0)\right) \left(\tan
   \left(\frac{1}{4} T_s (\omega_0+\omega_c)\right)+i\right)-2 i
   \omega_c \tan \left(\frac{T_s
   \omega_0}{2}\right)+\omega_0+\omega_c\right)}{\omega_0
   (\omega_0-\omega_c)
   (\omega_0+\omega_c)}+\frac{1}{\omega_0^2+\omega_0
   \omega_c}\right)  .	 
   \end{split}
\end{equation}

\noindent
It can be seen that the solutions obtained in Eqs. (\ref{y2t}) and (\ref{y3t}) are also equivalent to the Eqs. in (A8) of Ref. \cite{amico2}, in the limit of high frequency switching of the coupling $\varpi_s \rightarrow \infty$, $T_s \rightarrow 0$.

Therefore, truncating the perturbative expansion of the wavefunction at second order in terms of $\delta$ and considering only $n=0,1$ photon in the cavity, we obtain the following approximate solution of the Schr\"{o}dinger equation

\begin{eqnarray}
\label{psi_exp2q2}
\begin{split}
{\lvert \psi \left( t \right)  \rangle} =  {\lvert gg,0  \rangle}^{(0)} +   \frac{g_0 \delta} {2 (\omega_0+\omega_c)}\left(-1+\frac{2 e^{-i t
   (\omega_0+\omega_c)}}{1+e^{\frac{1}{2} i \frac{2 \pi}{\varpi_s} 
   (\omega_0+\omega_c)}}\right)
 \left[ {\lvert ge,1  \rangle}^{(1)} + {\lvert eg,1  \rangle}^{(1)} \right]  + \\ 
+  \frac{g_0^2 \delta^2 \left(i (2 t+\frac{2 \pi}{\varpi_s} ) (\omega_0+\omega_c)-2 i
   \left(1+e^{-i t (\omega_0+\omega_c)}\right) \tan
   \left(\frac{1}{4} \frac{2 \pi}{\varpi_s}  (\omega_0+\omega_c)\right)+2 e^{-i t
   (\omega_0+\omega_c)}-2 \sec ^2\left(\frac{1}{4} \frac{2 \pi}{\varpi_s} 
   (\omega_0+\omega_c)\right)\right)}{4 (\omega_0+\omega_c)^2} {\lvert gg,0  \rangle}^{(2)} +  \\
  + \frac{1}{4} g_0^2 \delta^2 \left(\frac{2 i e^{-i t
   (\omega_0+\omega_c)} \left(\tan \left(\frac{1}{4} \frac{2 \pi}{\varpi_s} 
   (\omega_0+\omega_c)\right)+i\right)}{\omega_0^2-\omega_c^2} + \right. \\  
   \left. +\frac{e^{-2 i t \omega_0} \left(2 \omega_0 \tan
   \left(\frac{1}{4} \frac{2 \pi}{\varpi_s}  (\omega_c-\omega_0)\right) \left(\tan
   \left(\frac{1}{4} \frac{2 \pi}{\varpi_s}  (\omega_0+\omega_c)\right)+i\right)-2 i
   \omega_c \tan \left(\frac{\frac{2 \pi}{\varpi_s}
   \omega_0}{2}\right)+\omega_0+\omega_c\right)}{\omega_0
   (\omega_0-\omega_c)
   (\omega_0+\omega_c)}+\frac{1}{\omega_0^2+\omega_0
   \omega_c}\right) {\lvert ee,0  \rangle}^{(2)}      .
\end{split}
\end{eqnarray}

\noindent
The result obtained in Eq. (\ref{psi_exp2q2}) does not depend on the particular definition of $\delta$ and $g_0$ as it depends on the product $\delta g_0$.
It is important to note that the wavefunction (\ref{psi_exp2q2}) has a parametric divergence for $\varpi_s \rightarrow 2 \omega_0$, $\varpi_s \rightarrow  \omega_0 + \omega_c$ and $\varpi_s \rightarrow  \omega_c - \omega_0$. For these values of the frequency of switching of the coupling the perturbative approach breaks down, as one of the contribution to the perturbative wavefunction becomes dominant over the others.



\end{document}